\documentclass[a4paper]{article}
\usepackage[latin1]{inputenc} 
\usepackage[T1]{fontenc} 
\usepackage{RR,RRthemes}
\usepackage{hyperref}

\usepackage{graphicx}
\usepackage{url}
\usepackage{subfigure}
\usepackage{xspace}

\newcommand{\ignore}[1]{}

\newcommand{\tabwidth}[0]{.33\linewidth}

\newcommand{\spacefig}[0]{{\vspace{-0.25in}\hspace{-1.2cm}}}
\newcommand{\spacefigtwo}[0]{{\vspace{-0.1in}\hspace{2.0cm}}}
\newcommand{\spacefigthree}[0]{{\vspace{-0.2in}\hspace{-1.2cm}}}
\newcommand{\tabbegin}[0]{\begin{tabular}{@{\spacefig}p{\tabwidth}p{\tabwidth}p{\tabwidth}@{}}}
\newcommand{\tabend}[0]{\\ \end{tabular}}

\newcommand{\tabbegintwo}[0]{\begin{tabular}{@{\spacefigtwo}p{\tabwidth}@{\spacefigtwo}p{\tabwidth}@{}}}
\newcommand{\tabendtwo}[0]{\\ \end{tabular}}

\newcommand{\tabbeginthree}[0]{\begin{tabular}{@{\spacefigthree}p{\tabwidth}p{\tabwidth}p{\tabwidth}@{}}}
\newcommand{\tabendthree}[0]{\\ \end{tabular}}

\newtheorem{theorem}{Theorem}[section]

\newtheorem{pb}[theorem]{Problem}
\newtheorem{prop}[theorem]{Property}

\RRdate{Février 2010}

\RRauthor{Emmanuel Agullo\thanks{HiePACS Team - INRIA Bordeaux Sud
    Ouest}%
  \and Jack Dongarra\thanks[utk]{University of Tennessee}%
  \and Rajib Nath\thanksref{utk}
  \and Stanimire Tomov\thanksref{utk}
}
\authorhead{Agullo, Dongarra, Nath \& Tomov}
\RRtitle{Optimisation automatique entièrement empirique pour la
  factorisation QR dense sur architectures multi-coeur}
\RRetitle{A Fully Empirical Autotuned Dense QR Factorization For
  Multicore Architectures}
\titlehead{Fully Empirical Autotuned QR Factorization For
  Multicore Architectures}
\RRresume{L'optimisation de librairies numériques est devenue de plus en plus
difficile, en même temps que les systèmes se sont complexifiées.  En
particulier, les machines multi-coeur modernes rendent le comportement
des algorithmes difficile à prévoir et modéliser.  Dans ce papier,
nous étudions le problème de l'optimisation d'une factorisation QR
dense sur des architectures multi-coeur.  Nous montrons qu'il est
difficile d'utiliser un modèle précis, ce qui nous motive pour
concevoir une méthode entièrement empirique. Nous mettons en avant
quelques propriétés empiriques vérifiées sur un large ensemble de
plate-formes. Ces propriétés nous permettent de réduire l'espace de
recherche. Notre méthode est automatique, rapide et fiable. Le
processus d'optimisation est en effet complètement effectué lors de
l'installation de la librairie en moins d'une heure et dix minutes
pour cinq des sept plate-formes étudiées. Nous atteigons une
performance moyenne variant de 97\% à 100\% de la performance optimale
selon les plate-formes. Ce travail est une base pour l'optimisation
automatique de la librairie PLASMA et permettre ainsi la portabilité
de sa performance.
}
\RRabstract{Tuning numerical libraries has become more difficult over time, as
systems get more sophisticated. In particular, modern multicore
machines make the behaviour of algorithms hard to forecast and
model. In this paper, we tackle the issue of tuning a dense QR
factorization on multicore architectures. We show that it is hard to
rely on a model, which motivates us to design a fully empirical
approach. We exhibit few strong empirical properties that enable us to
efficiently prune the search space.  Our method is automatic, fast and
reliable. The tuning process is indeed fully performed at install time
in less than one and ten minutes on five out of seven platforms. We
achieve an average performance varying from 97\% to 100\% of the
optimum performance depending on the platform. This work is a basis
for autotuning the PLASMA library and enabling easy performance
portability across hardware systems.
}
\RRmotcle{Optimisation automatique, optimisation empirique,
  multi-coeur, alg\`ebre lin\'eaire dense, factorisation QR} 

\RRkeyword{Autotuning, empirical tuning, multicore, dense linear algebra, QR
  factorization}
\RRprojets{HiePACS}
\RRdomaine{1} 
\RRthemeProj{hiepacs} 
\RCBordeaux 

\begin{document}
\RRNo{7526}
\makeRR   

\section{Introduction}
\label{sec:introduction}

The hardware trends have dramatically changed in the last few
years. The frequency of the processors has been stabilized or even
sometimes slightly decreased whereas the degree of parallelism has
increased at an 
exponential scale. This new hardware paradigm implies
that applications must be able to exploit parallelism at that same
exponential pace~\cite{sutterlunch}. Applications must also be able
to exploit a reduced bandwidth (per core) and a smaller amount of
memory (available per core). Numerical libraries, which are a critical
component in the stack of high-performance applications, must in
particular take advantage of the potential of these new
architectures. So long as library developers could depend on ever
increasing clock speeds and instruction level parallelism, they could
also settle for incremental improvements in the scalability of their
algorithms. But to deliver on the promise of tomorrow's petascale
systems, library designers must find methods and algorithms that can
effectively exploit levels of parallelism that are orders of magnitude
greater than most of today's systems offer.
Autotuning is therefore a major concern for the whole HPC community
and there exist many successful or on-going efforts. 
The FFTW library~\cite{FFTW98} uses autotuning techniques
to generate optimized libraries for FFT,
one of the most important techniques for digital signal processing.
Another successful example is the OSKI library~\cite{spmv} for
sparse matrix vector products.
The PetaBricks~\cite{petabricks} library
is a general purpose tuning method providing a language to describe
the problem to tune. It has several applications ranging from
efficient sorting~\cite{petabricks} to multigrid
optimization~\cite{petabricks-mg}.   
In the dense linear algebra community, several projects have tackled
this challenge on different hardware architectures. 
The Automatically
Tuned Linear Algebra Software (ATLAS) library~\cite{atlas} aims at
achieving high performance on a large range of CPU platforms
thanks to empirical tuning techniques performed at install time.
On graphic
processing units (GPUs), among others,~\cite{gputune-berk}
and~\cite{magma} have proposed efficient approaches. 
FLAME~\cite{flame09} and PLASMA~\cite{tileplasma} 
have been designed to achieve high performance on multicore
architectures thanks to tile algorithms (see
Section~\ref{sec:tileqr}).
The common characteristics of all these approaches are that they need
intensive tuning to fully benefit from the potential of the hardware.
Indeed, the increased degree of parallelism induces a more and more
complex memory hierarchy. 

Tuning a library consists of finding the parameters that maximize a
certain metric (most of the time the performance) on a given
environment. In general, the term \emph{parameter} has to be
considered in its broad meaning, possibly including a variant of an
algorithm. The \emph{search space}, corresponding to the possible set
of values of the \emph{tunable parameters} can be very large in
practice. Depending on the context, on the purpose and on the
complexity of the search space, different approaches may be
employed. Vendors can afford dedicated machines for delivering highly
tuned libraries~\cite{mkl,ibm,acml} and have thus limited constraints
in terms of time spent in exploring the search space. On the other
side of the spectrum, some libraries such as ATLAS aim at being portable and
efficient on a wider range of architectures and cannot afford a
virtually unlimited time for tuning. Indeed, empirical tuning is performed
at install time and there is thus a
trade-off between the time the user accepts to afford to install the
library and the quality of the tuning. In that case, the main
difficulty consists of efficiently pruning the search space. Of
course, once a platform has been tuned, the information can be shared
with the community so that it is not necessary to tune again the
library, but this is an orthogonal problem which we do not address
here. 
Model-driven tuning may allow one to efficiently prune the search
space. Such approaches have been successfully designed on GPU
architectures, in the case of matrix vector products~\cite{spmv} or dense linear
algebra kernels~\cite{gputune-berk,magma}. However, in practice,
the robustness of the assumptions on the model strongly depends both
on the algorithm to be tuned and on the target architecture. There is
no clearly identified trend yet but model-driven approaches
seem to be less robust on CPU architectures. For instance, even
in the single-core CPU case, basic linear algebra algorithms tend
to need more empirical search~\cite{atlas}. Indeed, on CPU-based
architectures, there are many parameters that are not under user
control and difficult to model (different levels of cache, different
cache policies at each level, possible memory contention, impact of
translation lookaside buffers (TLB) misses, \ldots) whereas the
current generations of GPU provide more control to the user.

In a previous work, we had tackled the issue of maximizing PLASMA
performance in order to compare it against other
libraries~\cite{perf}. We first manually pre-selected a combination of
parameters based on the performance of the most compute-intensive kernel. 
We then tried all these combinations for each
considered size of matrix to be factorized. This basic tuning approach
achieved high performance but required human intervention to
pre-select the parameters and days of run to find optimum
performance. In the present paper, not only we now tackle the issue of
automatically performing the tuning process but we also present new
heuristics that efficiently prune the search space so that the whole
tuning process is reduced to one hour or so. We illustrate our
discussion with the QR factorization implemented in the PLASMA
library, which is representative~\cite{perf} of all 
three one-sided factorizations
(QR, LU, Cholesky) currently available in PLASMA.
Because of the trends expose above and as further motivated
in Section~\ref{sec:motivation}, we do \emph{not} rely on a model
to tune our library.
Instead, we employ a fully empirical
approach and we exhibit few empirical properties that enable us to
efficiently prune the search space.

The rest of the paper is organized as
follows. Section~\ref{sec:problem} presents the problem and motivates
the outline of our two-step empirical
approach (Section~\ref{sec:outline}). Section~\ref{sec:env} presents
the wide range of hardware platforms used in the experiments to
validate our approach. Section~\ref{sec:step1} describes the first
empirical step, consisting of benchmarking the most compute-intensive
serial kernels. We propose three new heuristics that automatically
pre-select (PS) candidate values for the tunable
parameters. Section~\ref{sec:step2} presents the second empirical
step, consisting of benchmarking effective multicore QR
factorizations. We propose a new pruning approach, which we call
``prune as you go'' (PAYG), that enables to further prune the search
space and to drastically reduce the whole tuning process. We conclude
and present future work directions in Section~\ref{sec:conclusion}.

\section{Problem Description}
\label{sec:problem}

\subsection{Tile QR factorization}
\label{sec:tileqr}

The development of programming models that enforce asynchronous, out
of order scheduling of operations is the concept used as the basis for
the definition of a scalable yet highly efficient software framework
for computational linear algebra applications. In PLASMA, parallelism
is no longer hidden inside Basic Linear Algebra
Subprograms~(BLAS)~\cite{blasurl} but is brought to the fore to yield
much better performance. We do not present tile algorithms in details
(more details can be found~\cite{tileplasma}) but their principles.
The basic idea is to split the initial matrix of order $N$ into
$NT\times NT$ smaller square pieces of order $NB$, called \emph{tiles}. 
Assuming that $NB$ divides $N$, the equality $N=NT\times NB$
stands. The algorithms are then represented as a Directed Acyclic
Graph (DAG)~\cite{graph} where nodes represent tasks performed on
tiles, either panel factorization or update of a block-column, and
edges represent data dependencies among them. More details on tile
algorithms can be found~\cite{tileplasma}. PLASMA currently implements
three one-sided (QR, LU, Cholesky) tile factorizations. The DAG of the
Cholesky factorization is the least difficult to schedule since there
is relatively little work required on the critical path. LU and QR
factorizations have exactly the same dependency pattern between the
nodes of the DAG, exhibiting much more severe scheduling and numerical
(only for LU) constraints than the Cholesky factorization. Therefore,
tuning the QR factorization is somehow representative of the work to
be done for tuning the whole library. In the following, we focus on
the QR factorization of square matrices in double precision
statically scheduled in PLASMA.
\begin{figure}[tbph]
    \centering
    \subfigure[Panel factorization and corresponding updates.]{
    \label{fig:panel_qr}
    \includegraphics[scale=0.25]{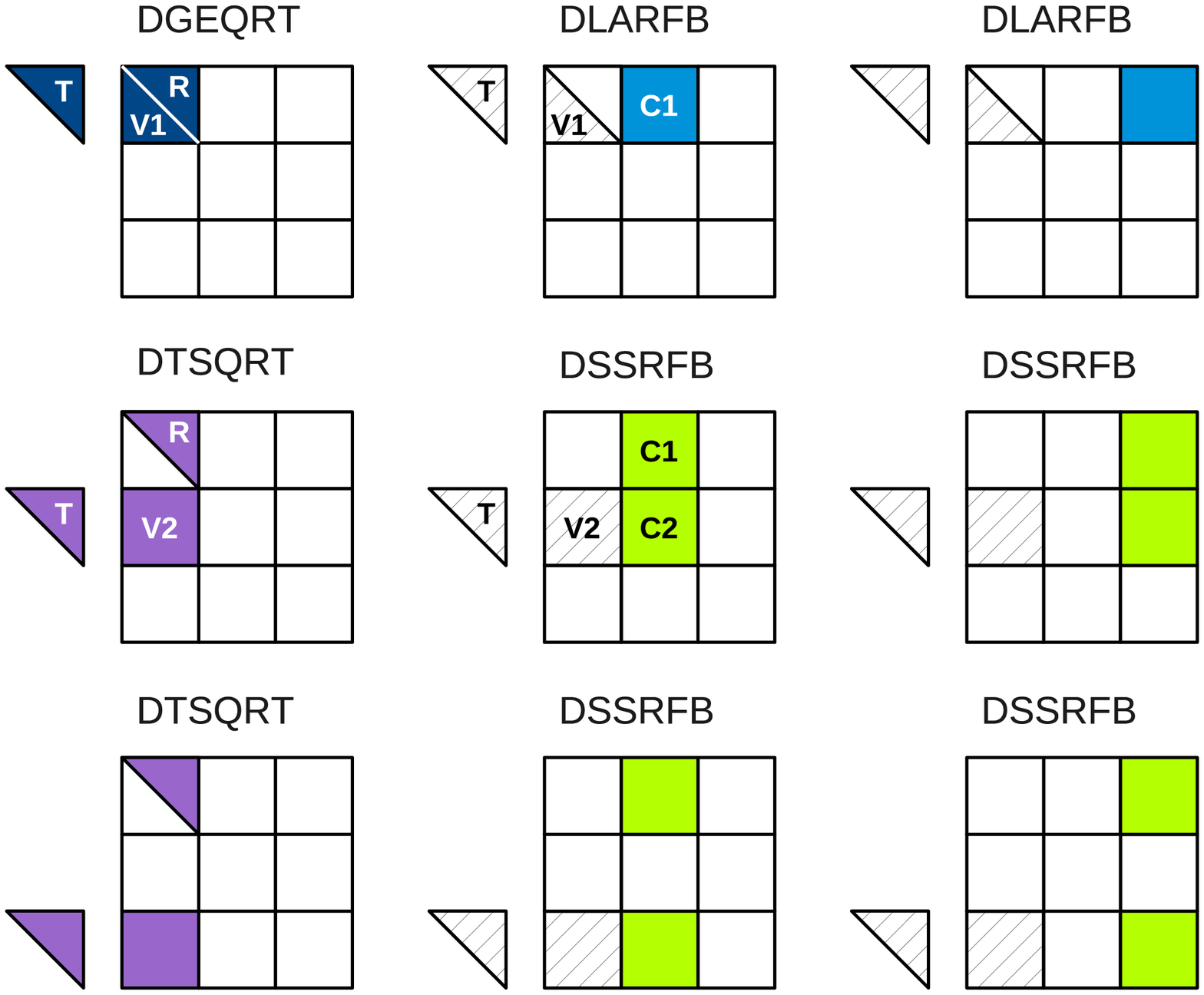}}
    \subfigure[DAG when the matrix is split in $5\times5$ tiles.]{
    \label{fig:dag_qr}
    \includegraphics[scale=0.20]{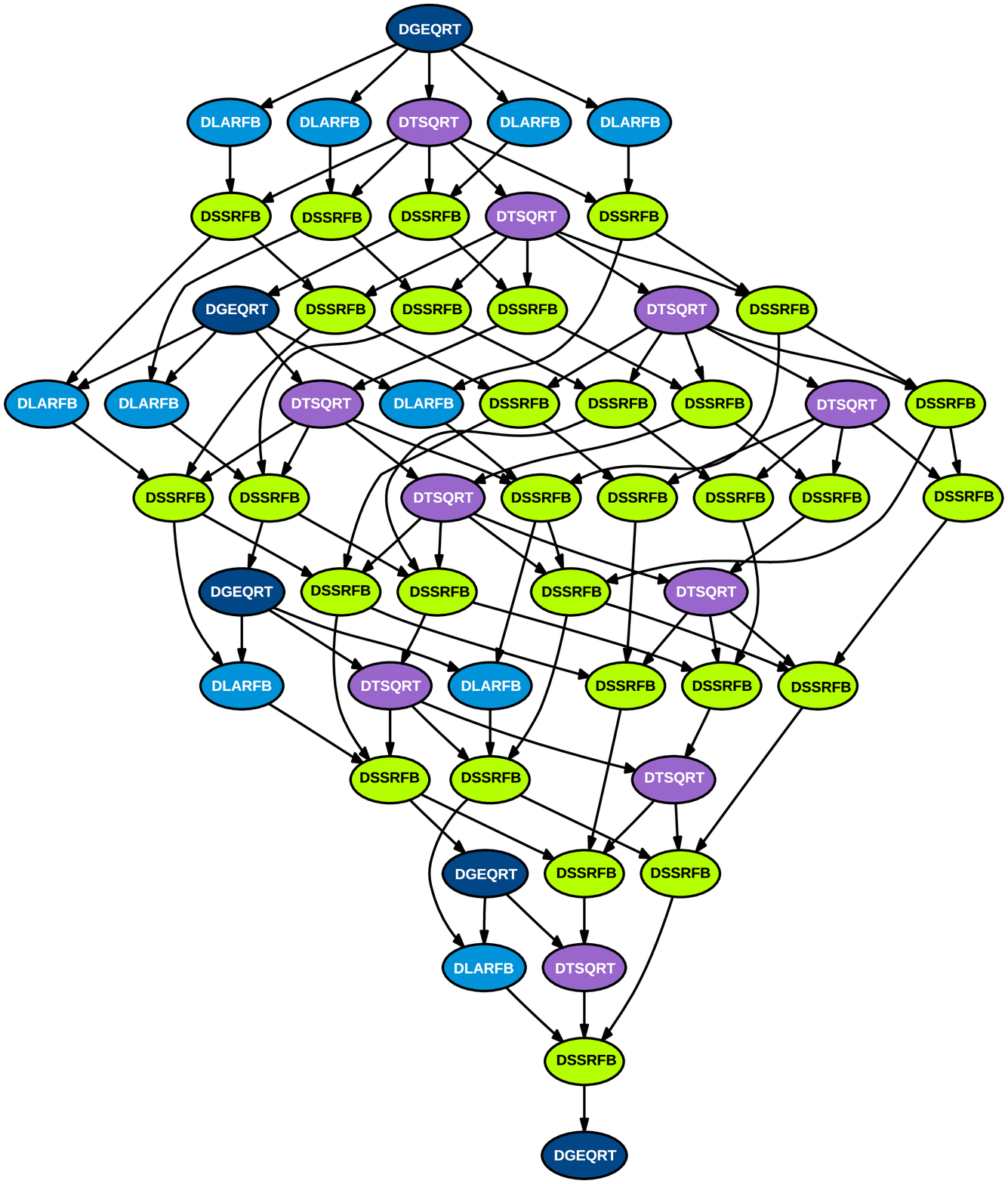}}
\caption{Tile QR Factorization}
\label{figure:panel_qr_dag_qr}
\end{figure}

Similarly to LAPACK which was built using a set of basic subroutines
(BLAS), PLASMA QR factorization is built on top of four serial
kernels. Each kernel indeed aims at being executed sequentially (by a
single core) and corresponds to an operation performed on one or a few
tiles. For instance, assuming a $3\times 3$ tile matrix,
Figure~\ref{fig:panel_qr} represents the first panel factorization
(DGEQRT and DTSQRT serial kernels~\cite{tileplasma}) and its
corresponding updates (DLARFB and DSSRFB serial
kernels~\cite{tileplasma}). The corresponding DAG (assuming this time
that the matrix is split in $5\times5$ tiles) is presented in
Figure~\ref{fig:dag_qr}.

\subsection{Tunable parameters and objective}

The shape of the DAG depends on the number of tiles ($NT\times NT$).
For a given matrix of order $N$, choosing the tile size NB is
equivalent to choosing the number of tiles (since $N=NB\times
NT$). Therefore, $NB$ is a first tunable parameter. A small value of
$NB$ induces a large number of tasks in the DAG and subsequently
enables the parallel processing of many tasks. On the other hand, the
serial kernel applied to the tiles needs a large enough granularity in
order to achieve a decent performance. The choice of NB thus trades
off the degree of parallelism with the efficiency of the serial
kernels applied to the tiles. There is a second tunable parameter,
called inner block size (IB). It trades off memory load with
extra-flops due to redundant calculations. With a value $IB=1$,
there are $\frac{4}{3}N^3$ operations as in standard LAPACK algorithm.
On the other hand, if no inner blocking
occurs ($IB=NB$), the resulting extra-flops overhead may 
represent $25\%$ of the whole QR factorization 
(see~\cite{tileplasma} for more details). 
The general objective of the paper is to address the following problem.
\begin{pb}
\label{pb}
Given a matrix size~$N$ and a number of cores~$ncores$, which tile
size and internal blocking size (NB-IB combination) do maximize the performance
of the tile QR factorization?
\end{pb}
Of course, the performance $P$ we aim at maximizing shall not depend
on extra-flops. Therefore, independently of the value of $IB$, 
we define $P=\frac{4}{3}\times N^3 / t$, where
$t$ is the elapsed time of the QR factorization. Note also that we
want the decision to be instantaneous when the user requests to
factorize a matrix so that the tuning process is to be performed
at install time.

\begin{figure}[tbph]
    \centering
    \subfigure[PLASMA QR factorization.]{
  \label{fig:core-1}
    \includegraphics[scale=0.45]{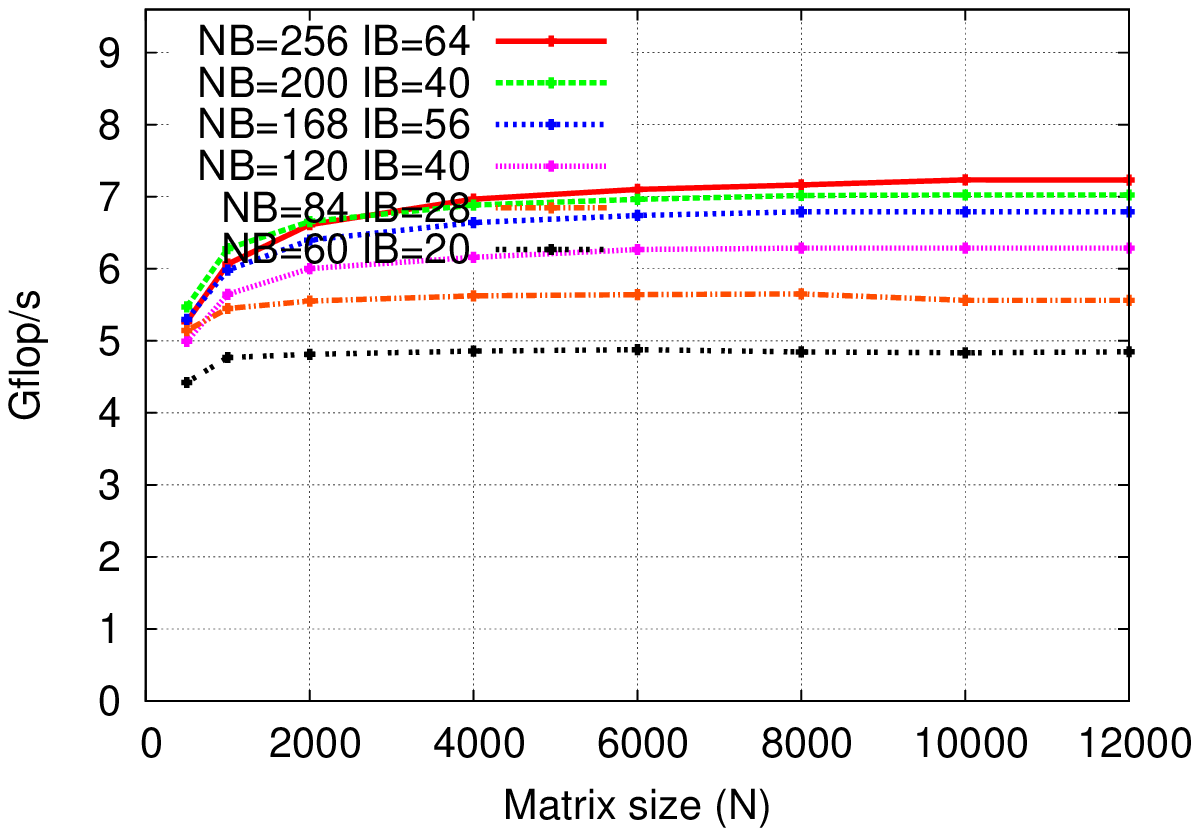}}
    \subfigure[Tile matrix multiplication (GEMM), with tile size $NB = 60$.]{
  \label{fig:gemm-nb60}
    \includegraphics[scale=0.45]{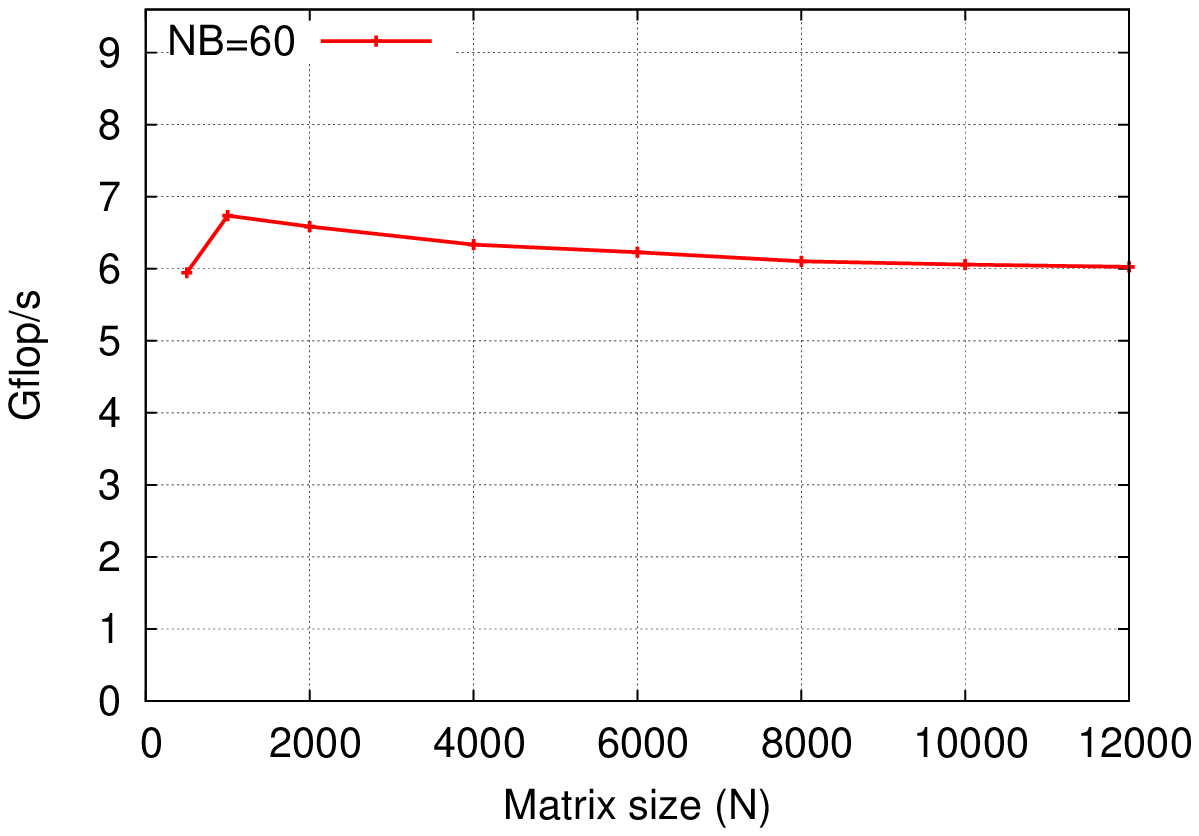}}
\caption{Performance of the PLASMA QR factorization (left) and tile
  matrix multiplication (right) on an Intel Core Tigerton machine.}
\label{fig:sequential}
\end{figure}
In a sequential execution of PLASMA, parallelism cannot be
exploited. In that case, PLASMA's performance is only related to
the performance of the serial kernel which increases with the tile
size. Figure~\ref{fig:core-1} illustrates this property on an Intel
Core Tigerton machine that will be described in details in
Section~\ref{sec:env}.

\begin{figure}[tbph]
    \centering
    \subfigure[Intel Tigerton machine - 16 cores.]{
  \label{fig:core-16}
    \includegraphics[scale=0.45]{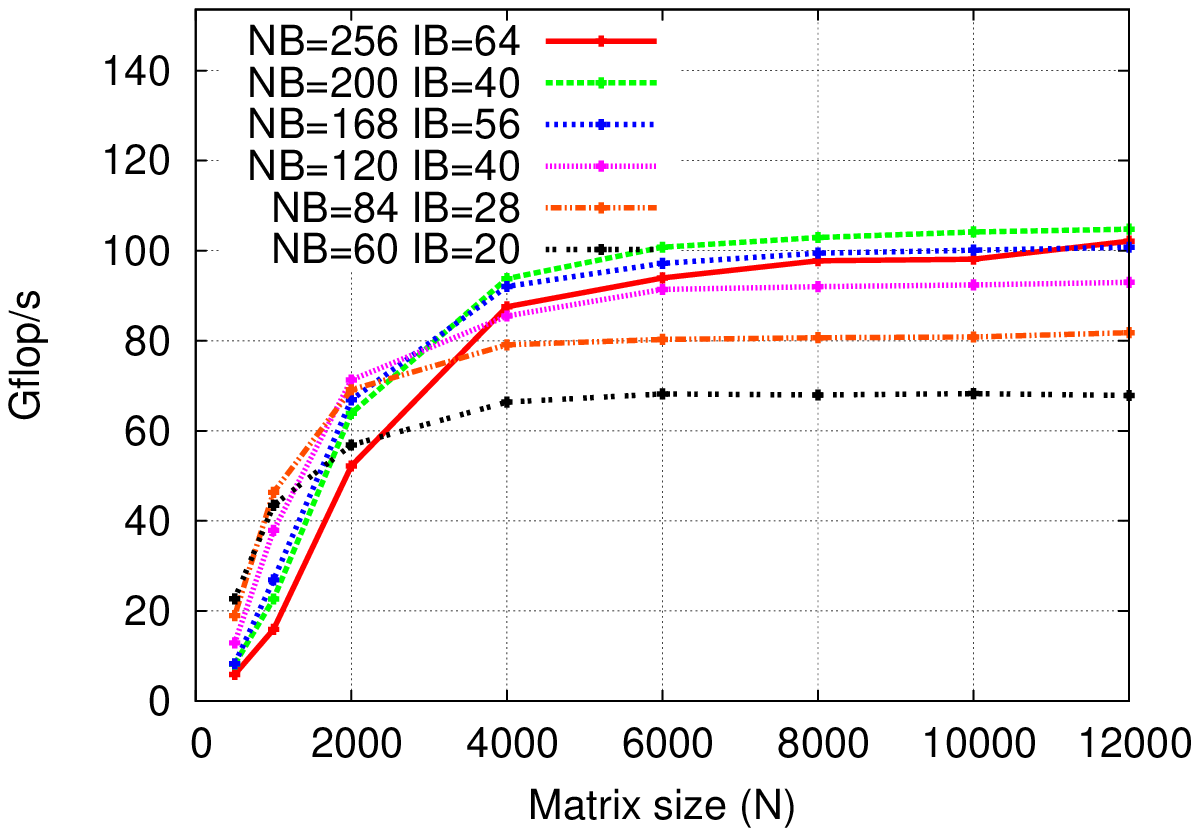}}
    \subfigure[IBM Power6 machine - 32 cores.]{
  \label{fig:power6-32}
    \includegraphics[scale=0.45]{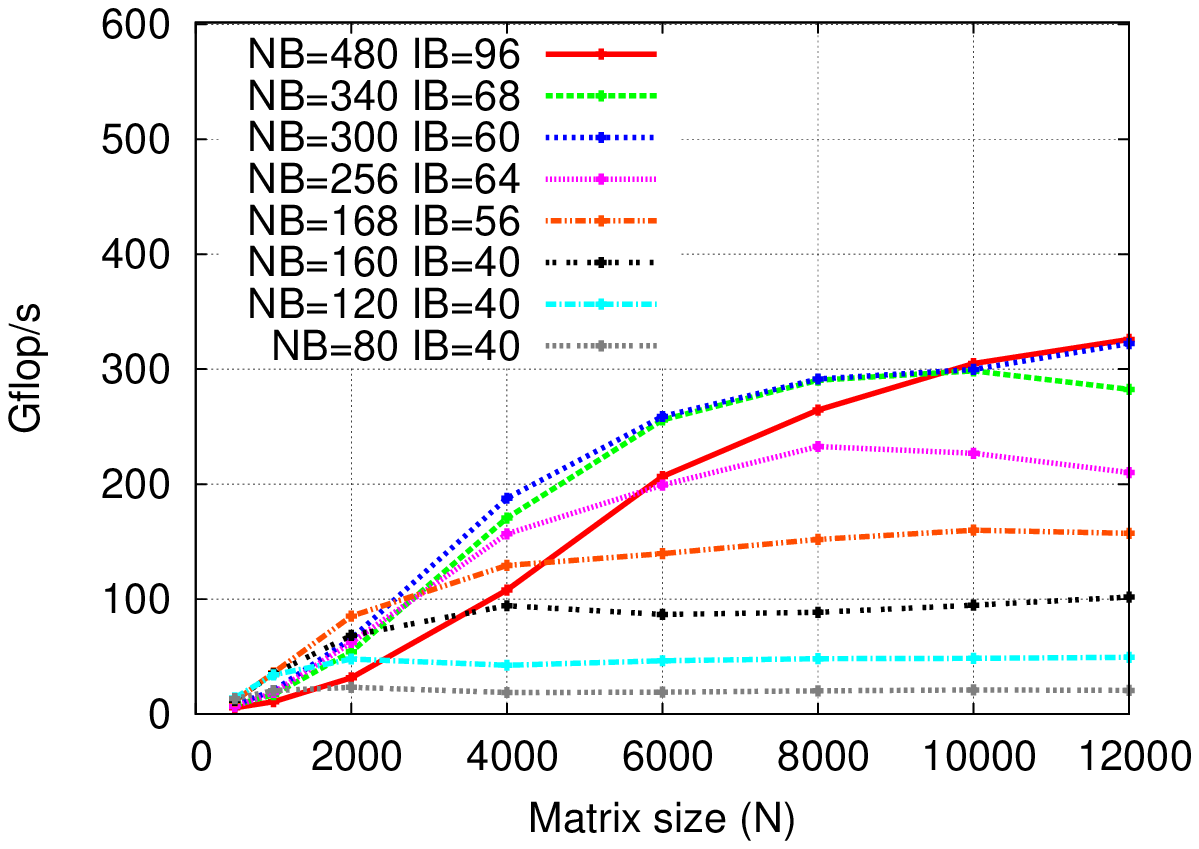}}
\caption{Performance of the PLASMA QR factorization}
\label{fig:core-16_power6-32}
\end{figure}
In a parallel execution of PLASMA, the optimum tile size depends on
the matrix size as shown on a $16$ cores execution in
Figure~\ref{fig:core-16}. Indeed, if the matrix is small, it needs to
be cut in even smaller pieces to provide work to all the $16$ 
cores even if this
induces that the serial kernels individually achieve a lower
performance. When the matrix size increases, all the cores may evenly
share the work using a larger tile size and thus achieving a higher
performance. In a nutshell, the optimum tile size both depends on the
number of cores and the matrix size, and its choice is critical for
performance.
Figure~\ref{fig:power6-32} shows that the impact is even stronger on a
32 cores IBM Power6 machine, also described in details in
Section~\ref{sec:env}.
The 80-40 combination is
optimum on a matrix of order $500$ but only achieves 
$6.3\%$ of the optimum ($20.6$ Gflop/s against
$325.9$ Gflop/s) on a matrix of order $12,000$.

\subsection{Motivation for an empirical approach}
\label{sec:motivation}

We have just shown the tremendous impact of the tunable parameters on
performance. As discussed in Section~\ref{sec:introduction}, the two
main classes of tuning methods are the model-driven and empirical
approaches. We mentioned in the introduction that dense linear algebra
algorithms are difficult to model on CPU-based architectures, and in
particular on multicore architectures. We now illustrate this
claim. Before coming back to the tile QR factorization, we temporarily
consider a simpler tile algorithm: the tile matrix multiplication: $C
\leftarrow C + A\times B$. Matrices $A$, $B$ and $C$ are split into
tiles $a_{ij}$, $b_{ij}$ and $c_{ij}$, respectively. The tile matrix
multiplication is then the standard nested loop on sub-arrays $i$, $j$
and $k$ whose single instruction is a DGEMM BLAS call on the
corresponding tiles: $c_{ij}\leftarrow c_{ij}+a_{ik}\times
b_{kj}$. Given the simplicity of this algorithm (simple DAG, only one
kernel, \ldots) one may expect that extrapolating the performance of
the whole tile algorithm $C \leftarrow C + A\times B$ from the
performance of the BLAS kernel $c_{ij}\leftarrow c_{ij}+a_{ik}\times
b_{kj}$ is trivial.

\begin{figure}[htbp]
    \centering
    \includegraphics[width=.6\linewidth]{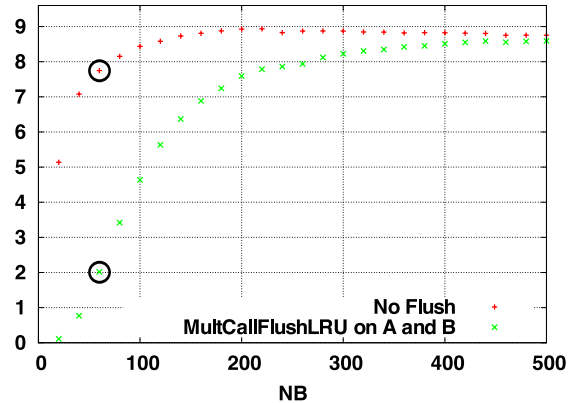}
    \caption{Performance (in Gflop/s) of a sequential matrix
      multiplication $c\leftarrow c+a\times b$ on the Intel Core
      Tigerton machine as a standard call to the vendor BLAS
      library. With the \emph{No Flush} strategy, data ($a$, $b$ and
      $c$) is \emph{not} flushed from the cache. With the
      \emph{MultCallFlushLRU} strategy~\cite{benchmarking}, $a$ and
      $b$ (but not $c$) are flushed from the cache. The values
      corresponding to a matrix order $NB=60$ are circled.}
  \label{fig:circles}
\end{figure}

However, the first difficulty is to correctly model how data are
accessed during the execution of the tile algorithms. Indeed, before
performing the BLAS call, some tiles may be in cache while others
are partially or fully out of cache. Figure~\ref{fig:circles} presents
the impact of the initial state of the tiles on the performance of a
sequential matrix multiplication $c\leftarrow c+a\times b$ on the
Intel Core Tigerton machine as a DGEMM call to the vendor BLAS
library. In the \emph{No Flush} strategy, all the tiles are initially
in cache (if they can fit). On the other hand, in the
\emph{MultCallFlushLRU}~\cite{benchmarking} strategy, $a$ and $b$ (but
not $c$) are flushed from the cache between two successive calls. To
achieve accurate timing, we called several times ($50$) the DGEMM
kernel for each matrix order ($NB$). The $50$ calls are timed all at
once; the average value finally computed is then more accurate than in
the case of timing a single call~\cite{benchmarking}. To simulate the
case where data is not flushed, all $50$ executions are performed on
the same data~\cite{benchmarking}. To simulate the case where $a$ and
$b$ are flushed, two large arrays $A$ and $B$ are allocated, and the
pointers $a$ and $b$ are moved along these arrays between two
successive calls. This self-flushing strategy was introduced
in~\cite{benchmarking}. Figure~\ref{fig:circles} shows that the impact
of the initial state is very important. For instance, for a tile of
order $NB=60$, the performance is four times higher ($8$ Gflop/s
against $2$ Gflop/s) in the \emph{No Flush} case. In practice, none of
these cases is a correct model for the kernel, since the sequential
tile multiplication based on a tile size $NB=60$ is neither $8$ nor
$2$ Gflop/s but $6$ Gflop/s as shown in Figure~\ref{fig:gemm-nb60}.

This experiment showed
that modeling tile algorithms on CPU-based architectures is not
trivial, even in the sequential case and even in the case of a simple
algorithm such as the matrix multiplication. Parallel execution
performance is even more difficult to forecast. For instance, frequent
concurrent accesses to the memory bus can slow down the memory
controller (as observed for small tile sizes on large matrices in
Figure~\ref{fig:power6-32}). The behavior of shared caches is also
difficult to anticipate. On top of that, other algorithmic factors
would add up to this complexity in the case of a more complex
operation such as a QR factorization. For instance, load
balancing issues and scheduling strategies 
must be taken into account when modeling a tile QR factorization.

As a consequence, we decided to base our approach on an extensive
empirical search coupled with only few but strongly reliable
properties to prune that search space.

\section{Two-step empirical method}
\label{sec:outline}

Given the considerations discussed in Section~\ref{sec:motivation},
we do \emph{not} propose a
model-driven tuning approach. Instead we use a fully empirical method
that effectively executes the factorizations on the target
platform. However, not all NB-IB combinations can be
explored. Indeed, an \emph{exhaustive search} is cumbersome since the
search space is huge. For instance, there are more than 1000 possible
NB-IB combinations even if we constrain NB to be an even integer
lower than 512 (size where the single core compute-intensive kernel
reaches its asymptotic performance) and if we impose IB to divide
NB. Exploring this search space on a matrix of order $N=10,000$ with 8
cores on the Intel Core Tigerton machine (described in
Section~\ref{sec:env}) would take several days. Therefore, we need to
prune the search space. We propose a two-step approach. In Step~1
(Section~\ref{sec:step1}), we benchmark the most compute-intensive
serial kernel. This step is fast since the serial kernels operate on
tiles, which are of small granularity ($NB<512$) compared to the matrices
to be factorized ($500\leq N\leq 10000$ in our study). Thanks to this
collected data set and a few well chosen empirical properties, we
pre-select (PS) a subset of NB-IB combinations. We propose three heuristics for
performing that preliminary pruning automatically. In step~2
(Section~\ref{sec:step2}) we benchmark the effective multicore QR
factorizations on the pre-selected set of NB-IB combinations. 
We furthermore show that further pruning (PAYG) can be performed 
during this step, drastically reducing the whole tuning process.

\section{Experimental environments}
\label{sec:env}


To assess the portability and reliability of our method, we consider
seven platforms based Intel EM64T processors, IBM Power and AMD
x86\_64. We recall here that we
are interested in shared memory multicore machines. Below is the list
of machines used in our experiments.

{\bf Intel Core Tigerton.} This 16 cores machine is a quad-socket
quad-core Xeon E7340 (codename Tigerton) processor, an Intel Core
micro-architecture. The processor operates at $2.39$ GHz. The
theoretical peak is equal to $9.6$ Gflop/s per core or $153.2$ Gflop/s
for the whole node, composed of 16 cores. There are two levels of
cache. The level-1 cache, local to the core, is divided into 32 kB of
instruction cache and 32 kB of data cache. Each quad-core processor
being actually composed of two dual-core Core2 architectures, the
level-2 cache has $2 \times 4$ MB per socket (each dual-core shares 4
MB). The effective bus speed is 1066 MHz per socket leading to a
bandwidth of $8.5$ GB/s (per socket).  The machine is running Linux
2.6.30 and provides Intel Compilers 11.0 together with the MKL 10.1
vendor library.

{\bf Intel Core Clovertown.} This 8 cores server is another machine
based on an Intel Core micro-architecture. The machine is composed of
two quad-core Xeon X5355 (codename Clovertown) processors, operating at
$2.66$ GHz.  The theoretical peak is equal to $10.64$ Gflop/s per core
and thus $85.12$ Gflop/s for the whole machine. The machine comes with
Linux 2.6.28, Intel Compilers 11.0 and MKL 10.1.

{\bf Intel Core Yorkfield.} This 4 cores desktop is also based on an
Intel Core micro-architecture. The machine is composed of one Core 2
Quad Q9300 (codename Yorkfield) processor, operating at $2.5$ GHz.
The theoretical peak is equal to $10.0$ Gflop/s per core and thus
$40.00$ Gflop/s for the whole machine with a shared 3 MB level-2 cache
per core pair. Each core has 64 KB of level-1 cache. The machine comes
with Linux 2.6.33, Intel Compilers 11.0 and MKL 10.1.

{\bf Intel Core Conroe.} This 2 cores desktop is based on an Intel
Core micro-architecture too. The machine is composed of one Core 2 Duo
E6550 (codename Conroe) processors, operating at $2.33$ GHz.  The
theoretical peak is equal to $9.32$ Gflop/s per core and thus $18.64$
Gflop/s for the whole machine with a shared 4 MB level-2 cache. Each
core has 128 KB of level-1 cache. The machine comes with Linux
2.6.30.3, Intel Compilers 11.1 and MKL 10.2.

{\bf Intel Nehalem.} This 8 cores machine is based on an Intel Nehalem
micro-architecture. Instead of having one bank of memory for all
processors as in the case of the Intel Core's architecture, each
Nehalem processor has its own memory. Nehalem is thus a Non Uniform
Memory Access (NUMA) architecture. Our machine is a dual-socket
quad-core Xeon X5570 (codename Gainestown) running at 2.93GHz and up
to 3.33 GHz in certain conditions (Intel Turbo Boost technology). The
Turbo Boost was activated during our experiments, allowing for a
theoretical peak of $13.32$ Gflop/s per core, i.e., $106.56$ Gflop/s
for the machine. Each socket has 8 MB of level-3 cache (that was
missing from most Intel Core-based microprocessors such as Tigerton
and Clovertown). Each core has 32 KB of level-1 instruction cache and
32 KB of level-1 data cache, as well as 256 KB of level-2 cache. The
machine comes with Linux 2.6.28, Intel Compilers 11.1 and MKL 10.2.

{\bf AMD Istanbul.} This 48 cores machine is composed of eight
hexa-core Opteron 8439 SE (codename Istanbul) processors running at
2.8 GHz. Each core has a theoretical peak of $11.2$ Gflop/s and the
whole machine $537.6$ Gflop/s. Like the Intel Nehalem, the Istanbul
micro-architecture is a ccNUMA architecture. Each socket has 6 MB of
level-3 cache. Each processor has a 512 KB level-2 cache and a 128 KB
level-1 cache. After having benchmarked the AMD ACML and Intel MKL
BLAS libraries, we selected MKL (10.2) which appeared to be slightly
faster in our experimental context. Linux 2.6.32 and Intel Compilers
11.1 were also used.

{\bf IBM Power6.} This 32 cores machine is composed of sixteen
dual-core IBM Power6 processors running at $4.7$ GHz. The theoretical
peak is equal to $18.8$ Gflop/s per core and $601.6$ Gflop/s for the
whole node. There are three levels of
cache. The level-1 cache, local to the core, can contain 64 kB of data
and 64 kB of instructions; the level-2 cache is composed of 4 MB per
core, accessible by the other core; and the level-3 cache is composed
of 32 MB common to both cores of a processor with one controller per
core (80 GB/s). The memory bus (75 GB/s) is shared by the 32 cores of
the node. The machine runs AIX 5.3 and provides the xlf 12.1 and xlc
10.1 compilers together with the Engineering Scientific Subroutine
Library (ESSL)~\cite{ibm} 4.3 vendor library.

\section{Step 1: Benchmarking the most compute-intensive serial kernel}
\label{sec:step1}

\begin{figure}[tbph]
    \centering
    \subfigure[Different NB-IB combinations with a
      common NB value have the same abscisse; ``Max IB'' represents
    the one that achieves the maximum performance among them.]{
    \includegraphics[scale=0.75]{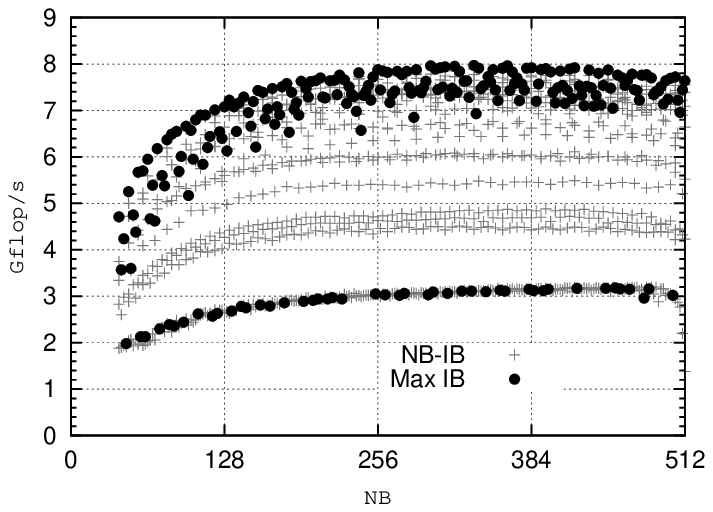}
    \label{fig:step1_a} 
    }
    \subfigure[Combinations pre-selected (PS) for each heuristic.]{
    \includegraphics[scale=0.75]{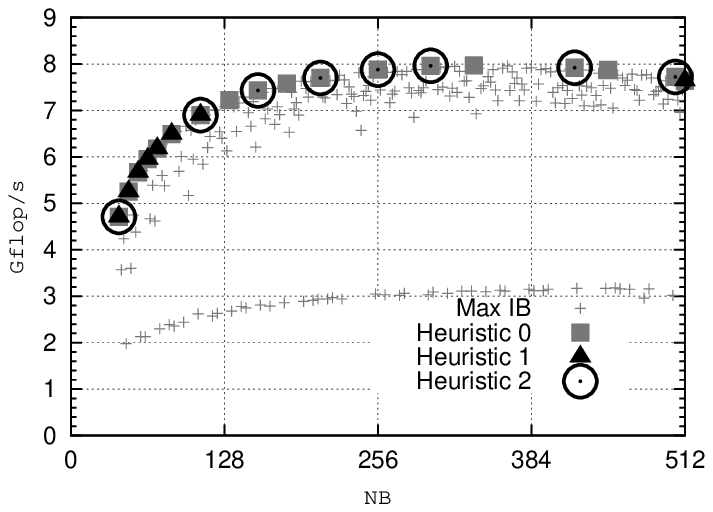}
    \label{fig:step1_b} 
    }
 \caption{Performance of the DSSRFB serial kernel depending
      on the NB-IB combination.} 
  \label{fig:step1}
\end{figure}

\begin{table}[htbp] 
\caption{Elapsed time (hh:mm:ss) for Step 1 and Step 2} 
\centering      
\begin{tabular}{l |r||c|| c|  c c}  
\hline\hline                        
\multicolumn{2}{c||}{Machine}&	Step 1           &                    \multicolumn{3}{c}{Step 2}\\
Architecture &	 \# cores&	           &		Heuristic&		PS	&	PSPAYG \\ [0.5ex]	
\hline                    
						&		&		&	0	&	14:46:37	&	03:05:41	\\
	Conroe	&	2	&	00:24:33	&	1	&	09:01:08	&	00:01:58	\\
						&		&
        &	2	&	07:30:53	&	{\bf 00:34:47}	\\
\hline                    
						&		&		&	0	&	17:40:00	&	04:48:13	\\
	Yorkfield	&	4	&	00:20:57	&	1	&	09:30:30	&	00:05:10	\\
						&		&
        &	2	&	08:01:05	&	{\bf 02:58:37}	\\
\hline                    
						&		&		&	0	&	20:08:43	&	02:56:25	\\
	Clovertown	&	8	&	00:21:44	&	1	&	11:06:18	&	00:13:09	\\
						&		&
        &	2	&	08:52:24	&	{\bf 01:10:53}	\\
\hline                    
						&		&		&	0	&	06:20:16	&	01:51:30	\\
	Nehalem	&	8		&	00:16:29	&	1	&	06:20:16	&	01:51:30	\\
						&		&
        &	2	&	06:20:16	&	{\bf 01:51:30}	\\
\hline                    
						&		&		&	0	&	23:29:35	&	03:15:41	\\
	Tigerton	&	16	&	00:34:18	&	1	&	12:22:06	&	00:08:57	\\
						&		&
        &	2	&	09:54:59	&	{\bf 01:01:06}	\\
\hline                    
						&		&		&	0	&	21:09:27	&	02:53:38	\\
    Istanbul	&	48	&	00:24:23	&	1	&	12:25:30	&	00:11:01	\\
						&		&
    &	2	&	10:04:46	&	{\bf 00:54:51}	\\
\hline                    
					&		&		&	0	& 03:06:05&	00:25:07\\
  Power6	&	32	&	00:15:23	&		1	& 03:06:05&	00:25:07\\
						&		&
  &	2	& 03:06:05&	{\bf 00:25:07}\\[1ex]
\hline     
\end{tabular} 
\label{table:step2_time}  
\end{table}
We explained in Section~\ref{sec:tileqr} that the tile QR
factorization consists of four serial kernels. However, the number of
calls to DSSRFB is proportional to $NT^3$ while the number of calls to
the other kernels is only proportional to $NT$ (DGEQRT) or to $NT^2$
(DTSQRT and DLARFB). Even on small DAGS (see Figure~\ref{fig:dag_qr}),
calls to DSSRFB are predominant. Therefore, the performance of this
compute-intensive kernel is crucial. DSSRFB's performance also depends
on NB-IB. It is thus natural to pre-select NB-IB pairs that allow
a good performance of DSSRFB before benchmarking the QR factorization
itself. The
practical advantage is that a kernel is applied at the granularity of
a tile, which we assume to be bounded by $512$ ($NB \leq
512$). Consequently, preliminary benchmarking this serial kernel can be
done exhaustively in a reasonable time. \emph{Step~1} 
thus consists of performing an exhaustive benchmarking of the
DSSRFB kernel on all possible NB-IB combinations and then to decide 
which of these will be kept for further testing in Step~2.
To achieve accurate timing, we followed the guidelines
of~\cite{benchmarking} as presented in
Section~\ref{sec:motivation}. In particular, DSSRFB is called $50$
times for each $(NB,IB)$ pair. We implemented both \emph{No Flush} and
\emph{MultCallFlushLRU} strategies. In this paper, we present results
related to the \emph{No Flush} approach. The reason is that it runs
faster and provides satisfactory results as we will show. A comparison
of both approaches is out of the scope of this manuscript.
Column
``Step 1'' of Table~\ref{table:step2_time} shows that the total
elapsed time for step 1 is acceptable on all the considered
architectures (between $16$ and $35$ minutes).
Figure~\ref{fig:step1_a} shows the resulting set of empirical
data collected during step 1 on the Intel Core Tigerton machine.
This data set can be pruned a first time. Indeed, contrary to NB,
which trades off parallelism for kernel performance, IB
only affects kernel performance but not parallelism. We can thus
perform the following orthogonal optimization:
\begin{prop}[Orthogonal pruning]
\label{prop:ib}
For a given NB value, we can safely pre-select the value of IB that
maximizes the kernel performance.
\end{prop}

Applying Property~\ref{prop:ib} to the data set of
Figure~\ref{fig:step1_a} results in discarding all NB-IB pairs except
the ones matching ``Max IB'', which still represents a large number of
combinations. We thus propose and assess three heuristics to further
prune the search space. The first considered heuristic is based on the
fact that a search performed with a well chosen subset of a limited
number - say 8 - of NB-IB combinations is enough to consistently achieve a
maximum performance for any matrix size $N$ or number of cores
$ncores$~\cite{perf}. Further intensive experiments led to
the following property.
\begin{prop}[Convex Hull]
\label{prop:lim}
There is consistently an optimum combination on the convex hull of the
data set.
\end{prop}
Therefore, {\bf Heuristic 0} consists of pre-selecting the
points from the convex hull of the data set (see
Figure~\ref{fig:step1_b}). In general, this approach may still
provide too many combinations. Because NB trades off kernel efficiency
with parallelism, the gains observed on kernel efficiency shall be
considered relatively to the increase of NB itself. Therefore, we
implemented {\bf Heuristic 1} that pre-selects the points of the
convex hull with a high steepness (or more accurately a point after a
segment with a high steepness). The drawback is that all these points
tend to be located in the same area as shown in
Figure~\ref{fig:step1_b} corresponding to small values of NB. To
correct this deficiency, we consider {\bf Heuristic 2} which first
divides the x-axis into iso-segments and pick up the point of maximum
steepness on each of these segments (see Figure~\ref{fig:step1_b}
again). Heuristics 1 and 2 are paremetrized to select a maximum of 8
combinations. All three heuristics perform a pre-selection (PS) 
that will be used as test cases for the second step. 

\section{Step 2: Benchmarking the whole QR factorization}
\label{sec:step2}

\subsection{Discretization and interpolation}
\label{sec:discretization}

\begin{figure}[htbp]
    \centering
    \subfigure[Intel Tigerton machine]{
    \includegraphics[scale=0.45]{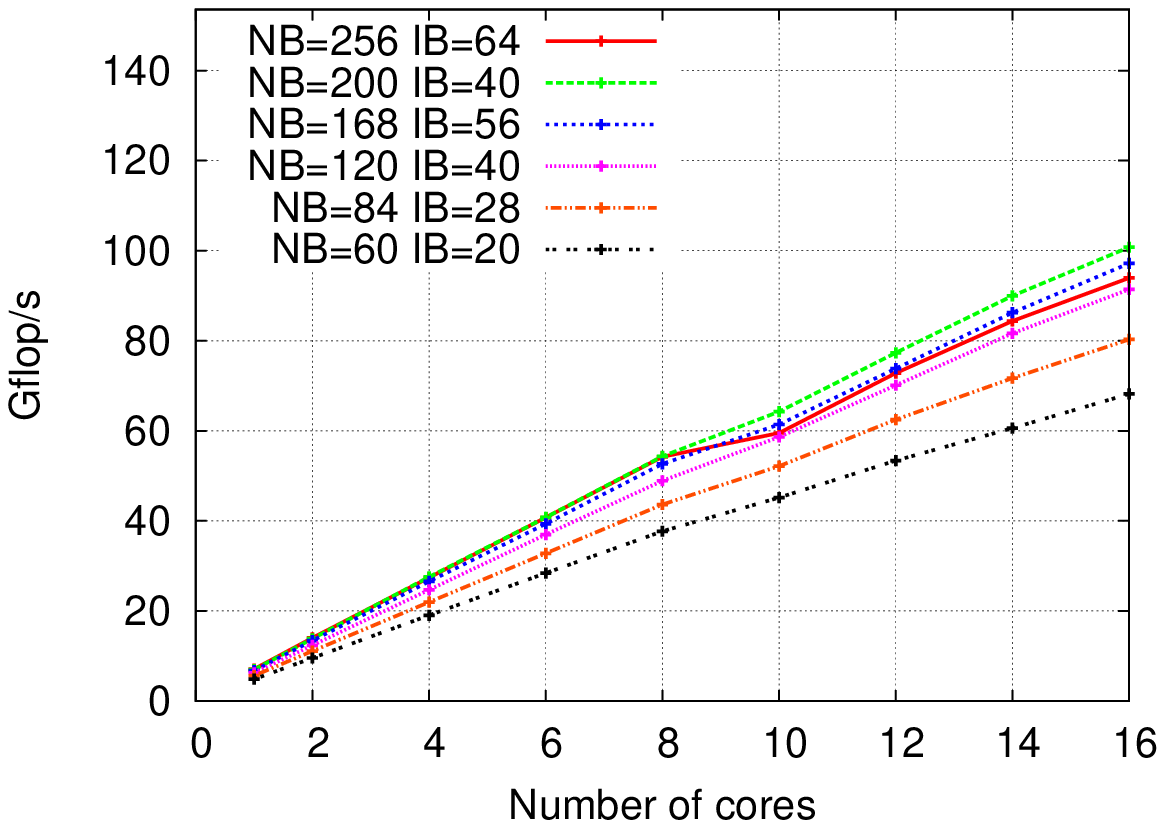}
  \label{fig:core-n6000}}
    \subfigure[IBM Power6 machine.]{
    \includegraphics[scale=0.45]{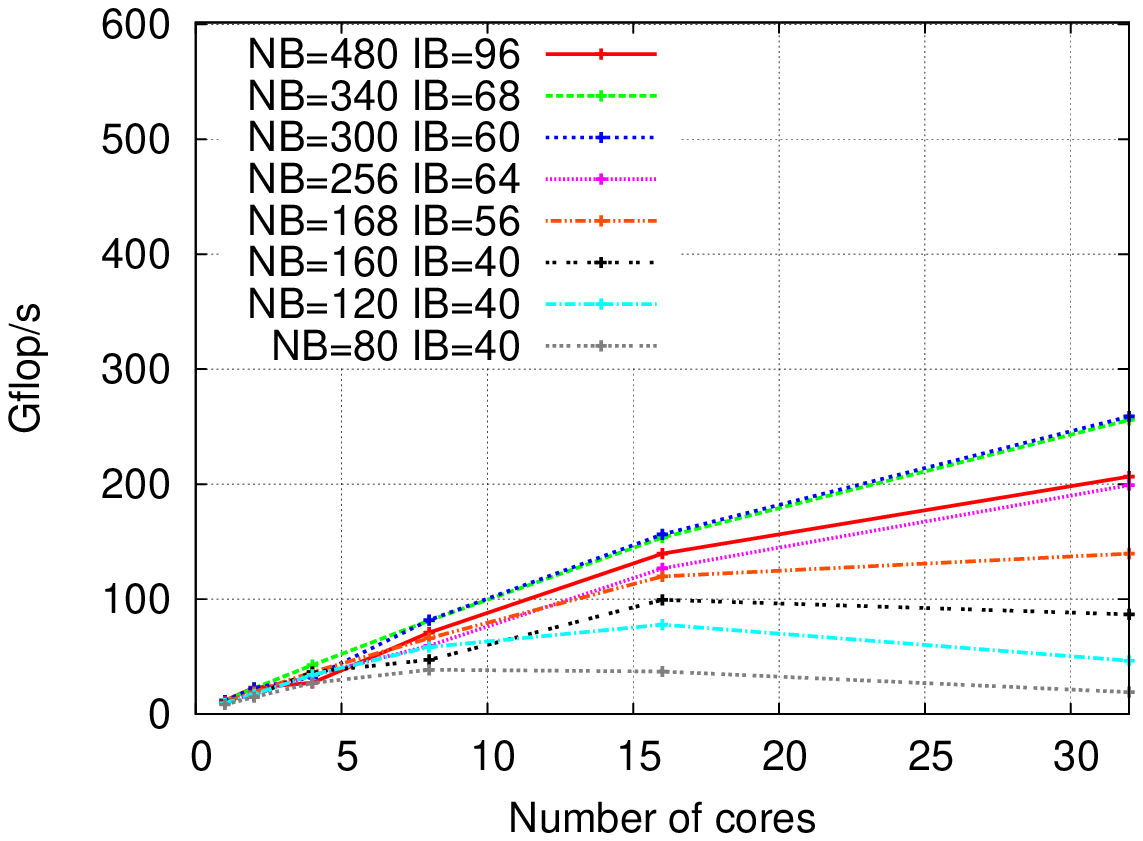}
  \label{fig:power6-n6000}}
\caption{Strong scalability - $N = 6000$.}
\label{fig:n6000}
\end{figure}

\begin{figure}[htbp]
    \centering
    \subfigure[Intel Tigerton machine]{
    \includegraphics[scale=0.45]{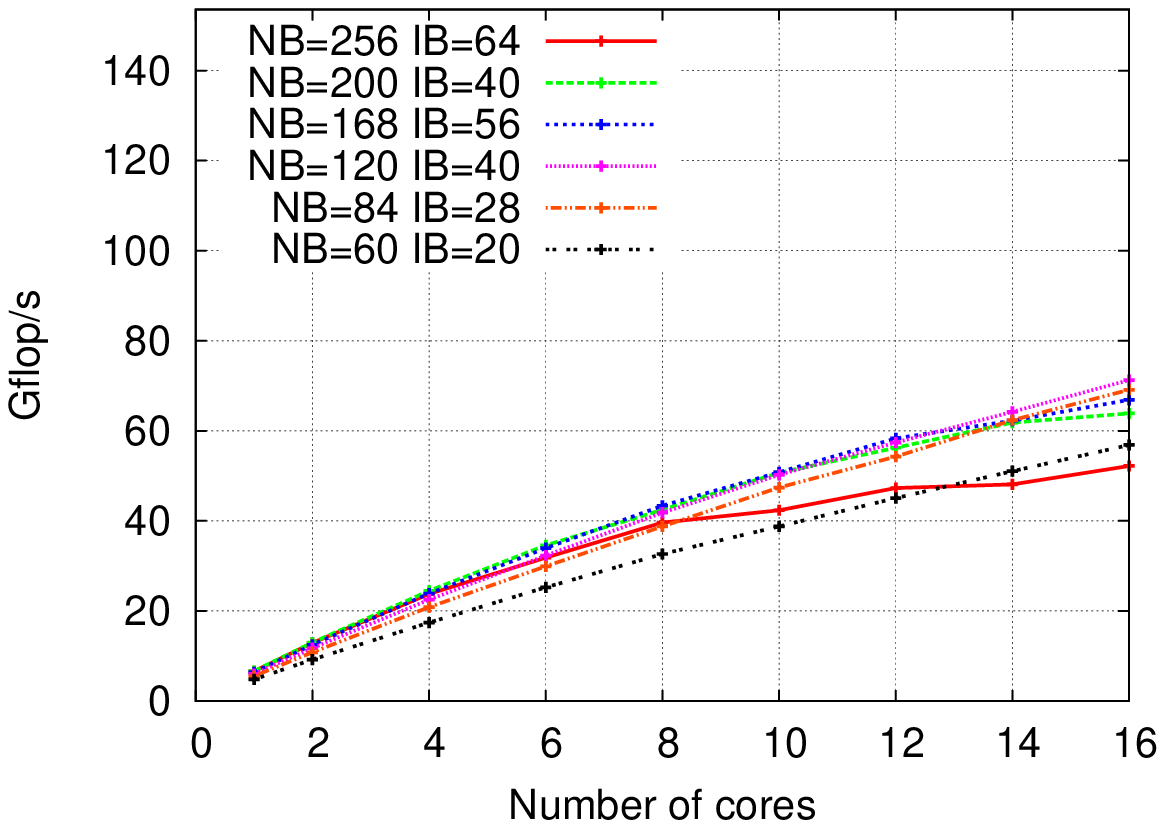}
  \label{fig:core-n2000}}
    \subfigure[IBM Power6 machine.]{
    \includegraphics[scale=0.45]{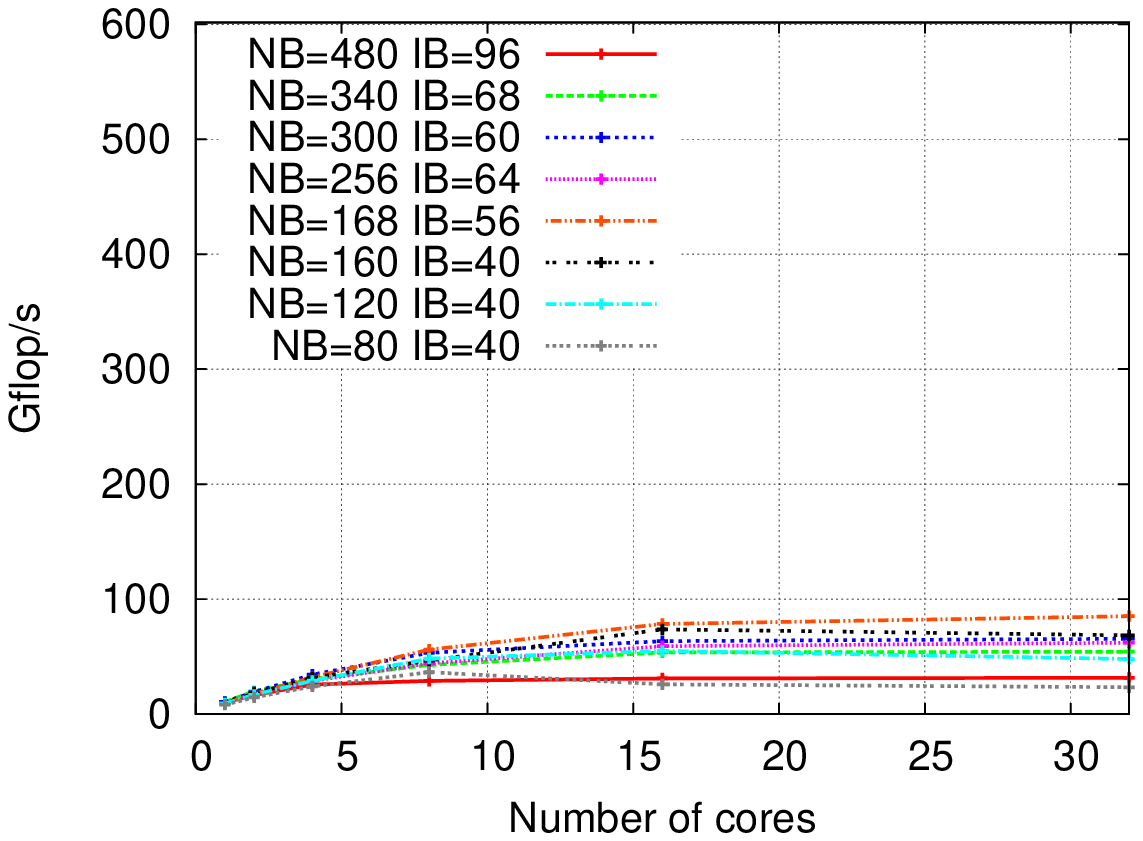}
  \label{fig:power6-n2000}}
\caption{Strong scalability - $N = 2000$.}
\label{fig:n2000}
\end{figure}

We recall that our objective is to immediately retrieve at execution
time the optimum NB-IB combination for the matrix size $N$ 
and number of cores $ncores$ that the user requests. Of course,
$N$ and $ncores$ are not known yet at install time. Therefore,
the ($N$,$ncores$) space to be benchmarked has to be discretized. 
We decided to benchmark all the
powers of two cores ($1$, $2$, $4$, $8$, \ldots) plus the maximum
number of cores in case it is not a power of two such as on the AMD
Istanbul machine. The motivation comes from empirical observation.
Indeed, Figures~\ref{fig:n6000} and~\ref{fig:n2000} show that the optimum
NB-IB combination can be finely interpolated with such a distribution. We
discretized more regularly the space on $N$ because the choice of the
optimum pair is much more sensible to that dimension (see
figures~\ref{fig:core-16} and~\ref{fig:power6-32}). We benchmarked
N=500, 1000, 2000, 4000, 6000, 8000, 10000\footnote{Except on the IBM
  Power6 machine where N=10000 was not benchmarked.}. Each run is
performed $6$ times to attenuate potential perturbations. When the
user requests the factorization of parameters that have not been tuned
(for instance N=1800 and ncores=5) we simply interpolate by selecting
the parameters of the closest configuration benchmarked at install
time (N=2000 and ncores=4 in that case).

\subsection{Impact of the pre-selection on the elapsed time of step~2}

Column PS (pre-selection) in Table~\ref{table:step2_time} shows the
impact of the heuristics (applied at step 1) on the time required for 
benchmarking step 2. Clearly Heuristic 0 induces a very long step 2 
(up to 1 day).
Heuristic 1 and 2 induce a lower time for step 2 (about $10$ hours)
but that may be still not acceptable for many users.

\subsection{Prune As You Go (PSPAYG)}

To further shorten step 2, we can perform complementary pruning on the
fly. Indeed, figures~\ref{fig:core-16} and~\ref{fig:power6-32} show
the following property.
\begin{prop}[Monotony]
\label{prop:cross}
Let us denote by $P(NB_1,N)$ and $P(NB_2,N)$ the performances obtained on a
matrix of order $N$ with tile sizes $NB_1$ and $NB_2$, respectively.
If $P(NB_1,N) > P(NB_2,N)$ and $NB_1 > NB_2$, then $P(NB_1,N') >
P(NB_2,N')$ for any $N'>N$.
\end{prop}
We perform step 2 in increasing order of $N$. After having benchmarked 
the current set of NB-IB combinations on a matrix of order $N$, 
we identify all the couples
$(NB_1,NB_2)$ that satisfy Property~\ref{prop:cross} and we remove
from the current subset the NB-IB pair in which $NB_2$ is involved.
Indeed, according to Property~\ref{prop:cross}, it would lead to a
lower performance than $NB_1$ on larger values of $N$ which are going
to be explored next. We denote this strategy by ``PSPAYG''
(pre-selection and prune as you go). Column PSPAYG in
Table~\ref{table:step2_time} shows that the time for step 2 is
dramatically improved with this technique. Indeed, the number of pairs
to explore decreases when $N$ increases, that is, when benchmark is
costly. For heuristic 2 (values in bold in
Table~\ref{table:step2_time}), the time required for step 2 is reduced
by a factor greater than $10$ in two cases (Intel Core Conroe and AMD
Istanbul machines).

\subsection{Reliability}

We employed the following methodology to assess the reliability of the
different tuning approaches. We first executed all the discussed
approaches on all the platforms with the discretization of the
($N$,$ncores$) space proposed in Section~\ref{sec:discretization}. We
then picked up between 8 and 16 ($N$,$ncores$) combinations such that
half of them were part of the discretized space (for instance $N=6000$
and $ncores=32$) and the other half were not part of it (for instance
$N=4200$ and $ncores=30$) so that the reliability of the interpolation
is also taken into account. For each combination we performed an
(almost) exhaustive search for reference. Detailed results are provided
in Appendix~\ref{sec:step2-details} and we now discuss the synthesis
of the results gathered in Table~\ref{table:val_conroe}. 
Heuristic 2 coupled with the
PSPAYG approach is very efficient since it achieves a high proportion
of the performance that would be obtained with an exhaustive search
(values in bold). The worst case occurs on the Istanbul machine, with
an average relative performance of $97.1\%$ (Column ``avg''). However,
even on that platform, the optimum NB-IB combination was found in
seven cases out of sixteen tests (Column ``optimum'').

Column $\frac{PSAYG}{PS}$ allows to specifically assess the impact of
the ``prune as you go'' method since they compare the average
performance obtained with PSPAYG (where pairs can be discarded during
step 2 according to Property~\ref{prop:cross}) compared to PS (where
no pair is discarded during step 2). The result is clear: pruning
during step 2 according to Property~\ref{prop:cross} does not hurt
performance ($\frac{|PS-PSPAYG|}{PS}<0.3\%$), showing that
Property~\ref{prop:cross} is strongly reliable. Finally, note that on
($N$,$ncores$) combinations part of the discretized space, PSPAYG
cannot achieve a higher performance than PS since all NB-IB
combinations tested with PSPAYG are also tested with PS. However,
PSPAYG can achieve a higher performance if ($N$,$ncores$) was not part
of the discretized space because of the interpolation. This is why
cases where $\frac{PSPAYG}{PS}>100\%$ may be observed.

\begin{table}[ht] 
\caption{Average performance achieved with a ``pre-selection'' (PS)
  method or a ``pre-selection and prune as you go'' (PSPAYG) method,
  based on different heuristics (H) applied at step 1. The performance
  is presented as a proportion of the exhaustive search (ES) or of the
  prunes search (PS). The column ``optimum'' indicates the number of
  times the optimum combination (with respect to the reference method)
  was found among the number of tests performed.} %
\centering      
\begin{tabular}{l| r| r r| r r| r r}  
\hline\hline                        
& & \multicolumn{2}{c}{$\frac{PS}{ES}$(\%)}\vline & \multicolumn{2}{c}{$\frac{PSPAYG}{ES}$(\%)}\vline & \multicolumn{2}{c}{$\frac{PSPAYG}{PS}(\%)$} \\
\hline                    
Machine & H & avg  & optimum &   avg  & optimum &  avg  & optimum \\ [0.5ex] 
\hline                    
	   &0 &99.67	&6/8	&99.67	&6/8	&100	&8/8\\
Conroe & 1 &95.28	&0/8	&95.28	&0/8	&100	&8/8\\
	   & 2 &99.54	&5/8	&{\bf 99.54}	&5/8	&100	&8/8\\
\hline                    
&0&98.63	&6/12	&98.63	&6/12	&100	&12/12\\
Yorkfield&1&91.53	&0/12	&91.59	&0/12	&100.07	&10/12\\
&2&98.63	&6/12	&{\bf 98.63}	&6/12	&100	&12/12\\
\hline                    
&0&98.59	&8/16	&98.35	&7/16	&99.76	&15/16\\
Clovertown&1&91.83	&0/16	&91.83	&0/16	&100	&16/16\\
&2&98.49	&9/16	&{\bf 98.25}	&8/16	&99.76	&15/16\\
\hline                    
&0&98.6	&8/16	&98.9	&8/16	&100.33	&16/16\\
Nehalem&1&98.6	&8/16	&98.9	&8/16	&100.33	&16/16\\
&2&98.6	&8/16	&{\bf 98.9}	&8/16	&100.33	&16/16\\
\hline                    
&0&97.36	&8/16	&97.54	&5/16	&100.21	&12/16\\
Tigerton&1&91.61	&0/16	&91.61	&0/16	&100	&16/16\\
&2&97.51	&8/16	&{\bf 97.79}	&7/16	&100.31	&15/16\\
\hline                    
&0&97.17	&7/16	&97.17	&7/16	&100	&16/16\\
Istanbul&1&94.12	&2/16	&94.12	&2/16	&100	&16/16\\
&2&97.23	&7/16	&{\bf 97.1}	&7/16	&99.87	&15/16\\
\hline                    
&0 &100        &16/16        &100        &16/16        &100        &16/16\\
Power 6&1 &100        &16/16        &100        &16/16        &100        &16/16\\
&2 &100        &16/16        &{\bf 100}        &16/16        &100        &16/16\\[1ex]

\hline     
\end{tabular} 
\label{table:val_conroe}  
\end{table}

\section{Conclusion}
\label{sec:conclusion}

We have presented a new fully autotuned method for dense linear algebra
libraries on multicore architectures. We have validated our approach
thanks to the PLASMA library on a wide range of architectures
representative of today's HPC CPU trends. We have illustrated our
discussion with the QR factorization, which is representative of the
difficulty of tuning any of the three one-sided factorizations (QR,
LU, Cholesky) present in PLASMA.

We have recalled that tuning consists in exploring a search space of
tunable parameters. In general, the exploration can be pruned thanks
to model-driven considerations. For our particular problem, we have
experimentally exhibited the weak reliability of the behavior of dense
linear algebra operations on modern multicore architectures. It
motivated us to use extensive empirical search coupled with only few
but strongly reliable properties to prune that search space. The
experimental validation has shown that the whole autotuning process
can in general be brought to completion in a decent time (less than
one hour and ten minutes on five out of seven platforms) 
though allowing to achieve a very high
performance (often finding the optimum tunable parameters and
achieving at least $97\%$ of the optimum performance in
average on each machine) on a wide range of architectures.

Our approach is user-friendly. At install time, the PLASMA library is
installed with default tunable parameters. A simple \emph{make
  autotune} launches the empirical benchmarking steps and builds the
decision tree based on simple interpolation properties. When the
end-user calls the library, the pre-built decision tree is used to
choose the optimized tunable parameters found at install time. The
process did not require any human intervention except on the IBM
platform. Indeed, we have had to manually arrange the tuning process
in order to cope with the batch scheduler (LoadLeveler~\cite{ll}) used
to submit the jobs on that IBM machine. We are currently working on a
better integration of the autotuning process with machines ruled by
batch schedulers.

We have considered the factorization of square matrices. The
factorization of non square matrices has to be studied too. In
particular, the case of tall and skinny matrices (which have a larger
number of rows than columns) often arises in several important
applications~\cite{demmel-tsqr}. In~\cite{tilecaqr}, the authors have
shown that communication-avoiding algorithms~\cite{demmel-tsqr} are
well adapted for processing such matrices in a multicore context.
They consist of splitting further the matrix in multiple block rows
(called domains) to enhance parallelism. The number $p$ of domains is
another tunable parameter that combines with NB and IB and should thus
be integrated in the empirical search method.

Hybrid multicore platforms with GPU accelerators tend to be more and
more frequent~\cite{plagma_qr}. The ultimate goal
being to develop a library that furthermore goes at scale when
increasing the number of nodes~\cite{dague}, the natural suite of the
work presented in that paper is to propose a unified framework for
tuning dense linear algebra libraries on modern hardware (distributed
memory, multicore micro-architectures and GPU accelerators). However,
the issues to be addressed, being very different from one
type of hardware to another, 
must
take into account the particularities of each type of hardware. In
that respect, the method presented in this document can be used as a
building block for such a unified framework.

\section*{Acknowledgment}
The authors would like to thank Jakub Kurzak, Greg Henry and Clint
Whaley for their constructive discussions.

\bibliographystyle{unsrt}
\bibliography{bibsc10}

\appendix
\section{Detailed results for Step 2}
\label{sec:step2-details}

We now present more detailed performance results to explain more
accurately how the synthetic results of Table~\ref{table:val_conroe}
were obtained. We illustrate our discussion with performance results
of the AMD Istanbul machine (tables~\ref{table:IG_ES},
\ref{table:IG_H0}, \ref{table:IG_H1} and~\ref{table:IG_H2}). To assess
the efficiency of the different methods presented in the paper, we
have performed between $8$ and $16$ tests on each machine. Each test
is an evaluation of the method for a given number of cores $ncores$
and a matrix size $N$. On the AMD Istanbul machine, the $16$ possible
combinations of N = 2000, 2700, 4200 or 6000 and $ncores$ = 4, 7, 40
or 48 have been tested. An exhaustive search (ES) is first performed
for all these $16$ combinations to be used as a reference
(Table~\ref{table:IG_ES}). Then we test which NB-IB combination would have been
chosen by the autotuner depending on the method it is built on
(tables~\ref{table:IG_H0}, \ref{table:IG_H1} and~\ref{table:IG_H2}).

We comment more specifically the results obtained for Heuristic~2
(Table~\ref{table:IG_H2}) since it is the heuristic that we plan to
set as a default in PLASMA. The first four rows show results related
to experimental conditions in which both the matrix order and the
number of cores are part of the values that were explicitly
benchmarked during the tuning process (N=2000 or 6000 and $ncores$=4
or 48). No interpolation is needed. In three cases, the optimum
configuration is found both by PS and PSPAYG. In the case were it was
not found (N=6000 and ncores=4) the optimum configuration was actually
not part of the initial pre-selected points by Heuristic 2 (Y=0). The
four next rows (N=2700 or 4200 and $ncores$=4 or 48) require to
interpolate the matrix order (but not the number of cores). For
N=2700, the selection is based on the benchmarking realized on
$N_0$=2000 while $N_0$=4000 is chosen when $N=4200$. The achieved
performance is not ideal since it is $8\%$ lower than the exhaustive
search. As expected, the interpolation on $ncores$ is much less
critical (four next rows). This observation confirms the validity of a
discretization coarser on the $ncores$ dimension. Finally (last four
rows), the quality of the tuning for the interpolation in both
dimensions is comparable to the one related to the interpolation on
$N$.
                                                  

\begin{table}[ht] 
\caption{Performance of ES on the AMD Istanbul Machine} 
\centering      
\begin{tabular}{r r r r r}  
\hline\hline                        
N & ncore & Perf (Gflop/s) & NB & IB \\ [0.5ex] 
\hline                    
2000 &	4  &	24.81  &	168	 & 28 \\
2000 &	48 &	140.1  &	96	 & 32 \\
6000 &	4  &	30.36  &	504	 & 56 \\
6000 &	48 &	272.55 &	168	 & 28 \\
\hline                    
2700 &	4  &	26.35  &	300	 & 60 \\
2700 &	48 &	176.7  &	108	 & 36 \\
4200 &	4  &	28.65  &	480	 & 60 \\
4200 &	48 &	239.93 &	128	 & 32 \\
\hline                    
2000 &	7  &	40.31  &	168	 & 28 \\
2000 &	40 &	135.72 &	96	 & 32 \\
6000 &	7  &	50.41  &	300	 & 60 \\
6000 &	40 &	236.8  &	168	 & 28 \\
\hline                    
2700 &	7  &	44.13  &	180	 & 36 \\
2700 &	40 &	168.79 &	108	 & 36 \\
4200 &	7  &	48.44  &	300  & 60 \\
4200 &	40 &	213.27 &	168  & 28 \\ [1ex]
\hline     
\end{tabular} 
\label{table:IG_ES}  
\end{table}

\setlength{\tabcolsep}{2pt}

\begin{table}[ht] 
\caption{Performance of Heuristic~0 on the AMD Istanbul machine.} 
\centering      
\begin{tabular}{r r r r r r r r } 
\hline\hline                        
N & ncore& Y &PS &	$\frac{PS}{ES}\%$	& PSPAYG & $\frac{PSPAYG}{ES}\%$	&$\frac{PSPAYG}{PS}\%$	\\ [0.5ex]
\hline                    
2000&  4&  1& 24.81&  100 	&24.81	&  100	&   100	 \\	
2000&  48& 1& 140.1&  100 &140.1	&  100	&   100	\\	
6000&  4&  1&30.36&  100 	&30.36	&  100	&   100	 \\	
6000&  48& 1& 272.55&  100	&272.55	&  100	&   100	 \\	
\hline                    
2700&  4&  1&24.24&  92	&24.24	&  92	&   100	 \\	
2700&  48& 1& 169.32&  95.83 &169.32	&  95.83&  	100	\\	
4200&  4&  1&26.8	 &  93.52&26.8	&  93.52&  	100	 \\	
4200&  48& 1& 237.19&  98.86&237.19	&  98.86&  	100	 \\	
\hline                    
2000&  7&  1&40.31&  100	&40.31	&  100	&   100	 \\	
2000&  40& 1& 126.66&  93.32&126.66	&  93.32&  	100	 \\	
6000&  7&  1&50.36&  99.9	&50.36	&  99.9	&   100	 \\	
6000&  40& 1& 236.8&  100	&236.8	&  100	&   100	 \\	
\hline                    
2700&  7&  1&40.4	 &  91.56 &40.4	&  91.56&  	100	 \\	
2700&  40& 1& 164.76&  97.61&164.76	&  97.61&  	100	 \\	
4200&  7&  1&44.64&  92.16&44.64	&  92.16&  	100	 \\	
4200&  40& 1& 213.27&  100	&213.27	&  100	&   100	 \\ [1ex]
\hline     
\end{tabular} 
\label{table:IG_H0}  
\end{table} 

\begin{table}[ht] 
\caption{Performance of Heuristic~1 on the AMD Istanbul machine} 
\centering      
\begin{tabular}{r r r r r r r r } 
\hline\hline                        
N & ncore& Y &PS &	$\frac{PS}{ES}\%$	& PSPAYG & $\frac{PSPAYG}{ES}\%$	&$\frac{PSPAYG}{PS}\%$	\\ [0.5ex]
\hline                    
2000	& 4	    & 0	& 22.84 & 92.06	& 22.84	& 92.06	& 100\\
2000	& 48    & 1 & 140.1 & 100	& 140.1	& 100	& 100\\
6000	& 4	    & 0	& 29.47	& 97.07	& 29.47	& 97.07	& 100\\
6000	& 48    & 0 & 256.42& 94.08	& 256.42& 94.08	& 100\\
\hline                    
2700	& 4	    & 0	& 22.9	& 86.92	& 22.9	& 86.92	& 100\\
2700	& 48    & 1 & 169.32& 95.83	& 169.32& 95.83	& 100\\
4200	& 4	    & 0	& 25.87	& 90.28	& 25.87	& 90.28	& 100\\
4200	& 48    & 1 & 239.93& 100	& 239.93& 100	& 100\\
\hline                    
2000	& 7	    & 0	& 36.92	& 91.57	& 36.92	& 91.57	& 100\\
2000	& 40    & 1 & 126.66& 93.32	& 126.66& 93.32	& 100\\
6000	& 7	    & 0	& 49.07 & 97.35	& 49.07	& 97.35	& 100\\
6000	& 40    & 0 & 224.13& 94.65	& 224.13& 94.65	& 100\\
\hline                    
2700	& 7	    & 0	& 38.83	& 88	& 38.83	& 88	& 100\\
2700	& 40	& 1 & 164.76& 97.61	& 164.76& 97.61	& 100\\
4200	& 7	    & 0	& 43.04	& 88.85	& 43.04	& 88.85 & 100\\
4200	& 40	& 0	& 209.74& 98.34	& 209.74& 98.34	& 100\\ [1ex]
\hline     
\end{tabular} 
\label{table:IG_H1}  
\end{table} 

\begin{table}[ht] 
\caption{Performance of Heuristic~2 on the AMD Istanbul machine} 
\centering      
\begin{tabular}{r r r r r r r r } 
\hline\hline                        
N & ncore& Y &PS &	$\frac{PS}{ES}\%$	& PSPAYG & \bf $\frac{PSPAYG}{ES}\%$	&$\frac{PSPAYG}{PS}\%$	\\ [0.5ex]
\hline                    
2000	&	4	&	1	&	24.81	&	100	&	24.81	&	100	&	100	\\
2000	&	48	&	1	&	140.1	&	100	&	140.1	&	100	&	100	\\
6000	&	4	&	0	&	29.98	&	98.75	&	29.35	&	96.66	&	97.89	\\
6000	&	48	&	1	&	272.55	&	100	&	272.55	&	100	&	100	\\
\hline                    
2700	&	4	&	1	&	24.24	&	92	&	24.24	&	92	&	100	\\
2700	&	48	&	0	&	169.32	&	95.83	&	169.32	&	95.83	&	100	\\
4200	&	4	&	1	&	26.8	&	93.52	&	26.8	&	93.52	&	100	\\
4200	&	48	&	0	&	237.19	&	98.86	&	237.19	&	98.86	&	100	\\
\hline                    
2000	&	7	&	1	&	40.31	&	100	&	40.31	&	100	&	100	\\
2000	&	40	&	1	&	135.72	&	100	&	135.72	&	100	&	100	\\
6000	&	7	&	1	&	50.36	&	99.9	&	50.36	&	99.9	&	100	\\
6000	&	40	&	1	&	236.8	&	100	&	236.8	&	100	&	100	\\
\hline                    
2700	&	7	&	0	&	40.4	&	91.56	&	40.4	&	91.56	&	100	\\
2700	&	40	&	0	&	157.06	&	93.05	&	157.06	&	93.05	&	100	\\
4200	&	7	&	1	&	44.64	&	92.16	&	44.64	&	92.16	&	100	\\
4200	&	40	&	1	&	213.27	&	100	&	213.27	&	100	&	100	\\ [1ex]
\hline     
\end{tabular} 
\label{table:IG_H2}  
\end{table}

\tableofcontents

\end{document}